\documentclass[twocolumn,showpacs,prl,aps,floatfix,reprint]{revtex4-1}
\usepackage{amssymb}
\usepackage{amsmath}
\usepackage{graphicx}
\usepackage{dcolumn}
\usepackage{amsfonts}
\usepackage{bm}
\newcommand{\bra}[1]{\left\langle #1\right|}
\newcommand{\ket}[1]{\left| #1\right\rangle}
\begin{document}

\title{Forging the link between nuclear reactions and nuclear structure}
\author{M. H. Mahzoon$^{1}$, R. J. Charity$^{2}$, W. H. Dickhoff$^{1}$, H. Dussan$^1$, S. J. Waldecker$^3$}
\affiliation{Department of Physics$^{1}$ and Chemistry$^{2}$, Washington University, St.
Louis, Missouri 63130}
\affiliation{Department of Physics$^3$, University of Tennessee, Chattanooga, TN 37403}
\date{\today}

\begin{abstract}
A comprehensive description of all single-particle properties associated with the nucleus ${}^{40}$Ca is generated by employing a nonlocal dispersive optical potential capable of simultaneously reproducing all relevant data above and below the Fermi energy.
The introduction of nonlocality in the absorptive potentials yields equivalent elastic differential cross sections as compared to local versions but changes the absorption profile as a function of angular momentum suggesting important consequences for the analysis of nuclear reactions.
Below the Fermi energy, nonlocality is essential to allow for an accurate representation of particle number and the nuclear charge density. Spectral properties implied by $(e,e'p)$ and $(p,2p)$ reactions are correctly incorporated, including the energy distribution of about 10\% high-momentum nucleons, as experimentally determined by data from Jefferson Lab.
These high-momentum nucleons provide a substantial contribution to the energy of the ground state, indicating a residual attractive contribution from higher-body interactions for ${}^{40}$Ca of about 0.64 MeV/$A$.
%Non-local optical potentials lead to substantial differences for the scattering wave functions in the interior of the %nucleus with corresponding consequences for the analysis of other nuclear reactions.

\end{abstract}

\pacs{21.10.Pc,24.10.Ht,11.55.Fv}
\maketitle

The properties of a nucleon that is strongly influenced by the presence of other nucleons have traditionally been studied in separate energy domains.
Positive energy nucleons are described by fitted optical potentials mostly in local form~\cite{Varner91,Koning03}.
Bound nucleons have been analyzed in static potentials that lead to an independent-particle model modified by the interaction between valence nucleons as in traditional shell-model calculations~\cite{Brown01,Caurier05}.
The link between nuclear reactions and nuclear structure is provided by considering these potentials as representing  different energy domains of one underlying nucleon self-energy.
This idea was implemented in the dispersive optical model (DOM) by Mahaux and Sartor~\cite{Mahaux91}.
By employing dispersion relations, the method provides a critical link between the physics above and below the Fermi energy with both sides being influenced by the absorptive potentials on the other side.

The DOM provides an ideal strategy to predict properties for exotic nuclei by utilizing extrapolations of these potentials towards the respective drip lines~\cite{Charity06,Charity13}.
The main stumbling block so far has been the need to utilize the approximate expressions for the properties of nucleons below the Fermi energy that were developed by Mahaux and Sartor~\cite{Mahaux91} to correct for the normalization-distorting energy dependence of the Hartree-Fock (HF) potential.
By restoring the proper treatment of nonlocality in the HF contribution, it was possible to overcome this problem~\cite{Dickhoff10} although the local treatment of the absorptive potentials yielded a poor description of the nuclear charge density and particle number.

In the present work we have for the first time treated the nonlocality of these potentials for the nucleus ${}^{40}$Ca with the aim to include \textit{all} available data below the Fermi energy that can be linked to the nucleon single-particle propagator~\cite{Dickhoff08} while maintaining a correct description of the elastic-scattering data. 
The result is a DOM potential that can be interpreted as the nucleon self-energy constrained by all available experimental data up to 200 MeV.
Such a self-energy allows for a consistent treatment of nuclear reactions that depend on distorted waves generated by optical potentials as well as overlap functions and their normalization for the addition and removal of nucleons to discrete final states. 
The re-analysis of such reactions may further improve the consistency of the extracted structure information.
Extending this version of the DOM to $N \ne Z$ will allow for predictions of properties that require the simultaneous knowledge of both reaction and structure information since at present few weakly-interacting probes are available for exotic nuclei~\cite{Dickhoff10a}. 

The self-energy $\Sigma_{\ell j}$ provides the critical ingredient to solve the Dyson equation for the nucleon propagator $G_{\ell j}$.  
Employing an angular momentum basis, it reads
\begin{eqnarray}
\label{eq:dyson}
G_{\ell j}(r,r';E) &=& G^{(0)}_{\ell j}(r,r';E) + \int \!\! d\tilde{r}\ \tilde{r}^2 \!\! \int \!\! d\tilde{r}'\ \tilde{r}'^2 \\
&\times&G^{(0)}_{\ell j}(r,\tilde{r};E)
\Sigma_{\ell j}(\tilde{r},\tilde{r}';E) G_{\ell j}(\tilde{r}',r';E) .
\nonumber
\end{eqnarray}
The noninteracting propagators $G^{(0)}_{\ell j}$ only contain kinetic energy contributions.
The solution of this equation generates $S_{\ell j}(r;E) =   \textrm{Im}\ G_{\ell j}(r,r;E)/\pi$, the hole spectral density, 
%\begin{equation}
%\label{eq:holes}
%\end{equation}
for negative continuum energies.
The spectral strength at $E$, for a given $\ell j$, is given by 
\begin{equation}
S_{\ell j}(E) = \int_{0}^\infty dr\ r^2\ S_{\ell j}(r;E) .
\label{eq:specs}
\end{equation}
For discrete energies one solves the eigenvalue equation for the overlap functions
%\begin{equation}
$\psi^n_{\ell j}(r) = \bra{\Psi^{A-1}_n}a_{r \ell j} \ket{\Psi^A_0}$,
%\label{eq:overlap}
%\end{equation}
%\begin{equation}
%\psi^n_{\ell j}(r) = \bra{\Psi^{A-1}_n}a_{r \ell j} \ket{\Psi^A_0} ,
%\label{eq:overlap}
%\end{equation}
for the removal of a nucleon at $r$ with discrete quantum numbers $\ell$ and $j$~\cite{Dickhoff10}.
The removal energy corresponds to
%\begin{equation}
$\varepsilon^-_n=E^A_0 -E^{A-1}_n$ 
%\label{eq:eig}
%\end{equation}
with the normalization for such a solution $\alpha_{qh}$ given by
%\begin{equation}
$S^n_{\ell j} = ( 1 - 
\partial \Sigma_{\ell j}(\alpha_{qh},
\alpha_{qh}; E)/\partial E |_{\varepsilon^-_n})^{-1} .
$
%\label{eq:sfac}
%\end{equation}
We note that from the solution of the Dyson equation below the Fermi energy, one can generate the one-body density matrix by integrating the non-diagonal imaginary part of the propagator up to the Fermi energy and therefore access the expectation values of one-body operators in the ground state including particle number, kinetic energy and charge density~\cite{Dickhoff08}. 
The latter is obtained by folding the point density with the nucleon form factors~\cite{Brown79}.
For positive energies, it was already realized long ago that the reducible self-energy provides the scattering amplitude for elastic nucleon scattering~\cite{Bell59}.

The self-energy fulfills the dispersion relation which relates the physics of bound nucleons to those that propagate at positive energy~\cite{Dickhoff08}.
It contains a
%\begin{eqnarray} 
%\mbox{Re}\ \Sigma_{\ell j}(r,r';E)\!& =& \! \Sigma^s_{\ell j} (r,r')\! - \! {\cal P} \!\!
%\int_{\varepsilon_T^+}^{\infty} \!\! \frac{dE'}{\pi} \frac{\mbox{Im}\ \Sigma_{\ell j}(r,r';E')}{E-E'}  \nonumber \\
%&+&{\cal P} \!\!
%\int_{-\infty}^{\varepsilon_T^-} \!\! \frac{dE'}{\pi} \frac{\mbox{Im}\ \Sigma_{\ell j}(r,r';E')}{E-E'} ,
%\label{eq:disprel}
%\end{eqnarray}
static correlated HF term and dynamic parts representing the coupling in the $A\pm1$ systems that start and end at the Fermi energies for addition ($\varepsilon_F^+ = E^{A+1}_0-E^A_0$) and removal ($\varepsilon_F^-=E^A_0-E^{A-1}_0$), respectively.
The latter feature is particular to a finite system and allows for discrete quasi particle and hole solutions of the Dyson equation where the imaginary part of the self-energy vanishes.
It is convenient to introduce the average Fermi energy
%\begin{equation}
$\varepsilon_F = \frac{1}{2} \left[
\varepsilon_F^+  - \varepsilon_F^- \right]$
%\label{eq:FE}
%\end{equation}
and employ the subtracted form of the dispersion relation calculated at this energy~\cite{Mahaux91,Dickhoff10}
\begin{eqnarray} 
\mbox{Re}\ \Sigma_{\ell j}(r,r';E)\! = \!  \Sigma_{\ell j} (r,r';\varepsilon_F)  \hspace{3.0cm} \label{eq:sdisprel} \\
- \! {\cal P} \!\!
\int_{\varepsilon_F^+}^{\infty} \!\! \frac{dE'}{\pi} \mbox{Im}\ \Sigma_{\ell j}(r,r';E') \left[ \frac{1}{E-E'}  - \frac{1}{\varepsilon_F -E'} \right]  \nonumber  \\
+{\cal P} \!\!
\int_{-\infty}^{\varepsilon_F^-} \!\! \frac{dE'}{\pi} \mbox{Im}\ \Sigma_{\ell j}(r,r';E') \left[ \frac{1}{E-E'}
-\frac{1}{\varepsilon_F -E'} \right]  ,
\nonumber
\end{eqnarray}
where $\mathcal{P}$ represents the principal value.
The beauty of this representation was recognized by Mahaux and Sartor~\cite{Mahaux86,Mahaux91} since it allows for a link with empirical information both for the real part of the nonlocal self-energy at the Fermi energy (probed by a multitude of HF calculations) as well as through empirical knowledge of the imaginary part of the optical potential also constrained by experimental data.
Consequently Eq.~(\ref{eq:sdisprel}) yields a dynamic contribution to the real part linking \textrm{both} energy domains around the Fermi energy.
Empirical information near $\varepsilon_F$ is emphasized by Eq.~(\ref{eq:sdisprel}) because of the $E'^{-2}$-weighting in the integrands.
The real self-energy at the Fermi energy will be denoted in the following by $\Sigma_{HF}$.

We now provide a more detailed description of the changes that are necessary in the conventional application of the DOM in order for the resulting potential to yield a realistic description of the single-particle properties below the Fermi energy.
In particular we refer to previous papers for a description of ingredients that have not changed from the purely local treatment of the DOM~\cite{Charity07,Mueller11}.
The nonlocal treatment of the HF potential was discussed in Ref.~\cite{Dickhoff10}.
The present form reads
\begin{eqnarray}
\Sigma_{HF}\left( \bm{r},\bm{r}' \right)   =   -V_{HF}^{vol}\left( \bm{r},\bm{r}'\right) 
+ V_{HF}^{wb}(\bm{r},\bm{r}') ,
\label{eq:HF}
\end{eqnarray}
where the volume term is given by
\begin{eqnarray}
V_{HF}^{vol}\left( \bm{r},\bm{r}' \right) = V_{HF}^0
\,f \left ( \tilde{r},r^{HF},a^{HF} \right ) \hspace{1.0cm}  \label{eq:HFvol} \\ 
\times \left [ x_1 H \left( \bm{s};\beta_{vol_1} \right) + (1-x_1) H \left( \bm{s};\beta_{vol_2}\right) \right ] ,
\nonumber
\end{eqnarray}
allowing for two different nonlocalities with different weight ($0 \le x_1 \le1$). 
We use the notation $\tilde{r} =(r+r')/2$ and $\bm{s}=\bm{r}-\bm{r}'$.
A wine bottle ($wb$) shape producing Gaussian is introduced replacing the surface term of Ref.~\cite{Mueller11}
\begin{equation}
V_{HF}^{wb}(\bm{r},\bm{r}') = V_{wb}^0  \exp{\left(- \tilde{r}^2/\rho_{wb}^2\right)} H \left( \bm{s};\beta_{wb} \right ).
\end{equation}
This Gaussian centered at the origin helps to represent overlap functions generated by simple potentials that reproduce corresponding Monte Carlo results~\cite{Brida11}. 
Non-locality is represented by a Gaussian form
\begin{equation}
H \left( \bm{s}; \beta \right) = \exp \left( - \bm{s}^2 / \beta^2 \right)/ (\pi^{3/2} \beta^3)
\end{equation}
first suggested in Ref.~\cite{Perey62}.
As usual we employ Woods-Saxon form factors $f(r,r_{i},a_{i})=[1+\exp \left({\frac{r-r_{i}A^{1/3}}{a_{i}}%
}\right)]^{-1}$.
Equation~(\ref{eq:HF}) is supplemented by the Coulomb and local spin-orbit interaction as in Ref.~\cite{Mueller11}.

The introduction of nonlocality in the imaginary part of the self-energy is well-founded theoretically both for long-range correlations~\cite{Waldecker2011} as well as short-range ones~\cite{Dussan11}. 
Its implied $\ell$-dependence is essential in reproducing the correct particle number for protons and neutrons.
The nonlocal part of this imaginary component has the form
%\begin{eqnarray}
%\textrm{Im}\ \Sigma( \bm{r},\bm{r}',E) =  \hspace{5.5cm} \label{eq:imnl} \\
%-W^{vol}_0(E) f\left(\tilde{r};r^{vol};a^{vol}\right)H \left( \bm{s}; \beta_{vol}\right) \hspace{1.0cm}
%\nonumber
%\\
%+4a^{sur}W^{sur}\left( E\right)H \left( \bm{s}; \beta_{sur}\right) \frac{d}{d \tilde{r} }f(\tilde{r},r^{sur},a^{sur}) .
%\nonumber
%\end{eqnarray}
\begin{eqnarray}
\textrm{Im}\ \Sigma( \bm{r},\bm{r}',E) =   
-W^{vol}_0(E) f\left(\tilde{r};r^{vol};a^{vol}\right)H \left( \bm{s}; \beta_{vol}\right)  
\nonumber
\\
+4a^{sur}W^{sur}\left( E\right)H \left( \bm{s}; \beta_{sur}\right) \frac{d}{d \tilde{r} }f(\tilde{r},r^{sur},a^{sur}) .
\hspace{0.5cm}
\label{eq:imnl}
\end{eqnarray}
%with the derivative taken at $\tilde{r} = \left(r+r'\right)/2$. 
We also include a local spin-orbit contribution as in Ref.~\cite{Mueller11}.
The energy dependence of the volume absorption has the form used in Ref.~\cite{Mueller11} whereas for surface absorption we employed the form of Ref.~\cite{Charity07}.
The solution of the Dyson equation below the Fermi energy was introduced in Ref.~\cite{Dickhoff10}.
The scattering wave functions are generated with the iterative procedure outlined in Ref.~\cite{Michel09} leading to a modest increase in computer time as compared to the use of purely local potentials.
Neutron and proton potentials are kept identical in the fit except for the Coulomb potential for protons.
The numerical values of all parameters together with a complete list of all employed equations is available in Ref.~\cite{Hossein13}.

Included in the present fit are the same elastic scattering data and level information considered in Ref.~\cite{Mueller11}.
%We refer to that paper for references to these data.
In addition, we now include the charge density of ${}^{40}$Ca as given in Ref.~\cite{deVries1987} by a sum of Gaussians in the fit. 
Data from the $(e,e'p)$ reaction at high missing energy and momentum obtained at Jefferson Lab for ${}^{12}$C~\cite{Rohe04}, ${}^{27}$Al, ${}^{56}$Fe, and ${}^{197}$Au~\cite{Rohe04A} were incorporated as well.
We note that the spectral function of high-momentum protons per proton number is essentially identical for ${}^{27}$Al and ${}^{56}$Fe thereby providing a sensible benchmark for their presence in ${}^{40}$Ca. 
We merely aimed for a reasonable representation of these cross sections since their interpretation requires further consideration of rescattering contributions~\cite{Barbieri06}.
%A detailed representation of these data would require our potentials to exhibit a energy-dependent geometry which %would increase the computational effort exponentially as discussed in Ref.~\cite{Dickhoff10}.
We did not include the results of the analysis of the $(e,e'p)$ reaction from NIKHEF~\cite{Kramer89} because the extracted  spectroscopic factors depend on the employed local optical potentials. We plan to reanalyze these data with our nonlocal potentials in a future study.

\begin{figure}[tbp]
\includegraphics*[scale=0.4]{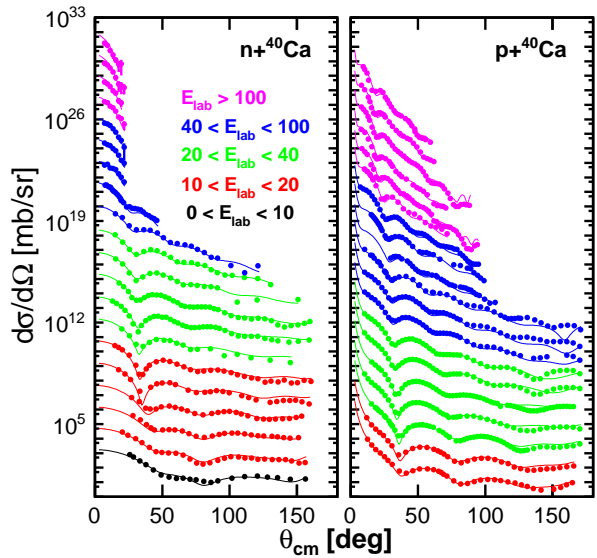}
\caption{(Color online) Calculated and experimental elastic-scattering
angular distributions of the differential cross section $d\protect\sigma %
/d\Omega $. Panels shows results for \textit{n}+$%
^{40}$Ca and \textit{p}+ $^{40}$Ca. Data for each energy are offset for clarity with the lowest energy at the bottom and highest at the top of each frame. References to the data are given in Ref.~\protect\cite{Mueller11}.}
\label{fig:elast}
\end{figure}
Motivated by the work of Refs.~\cite{Waldecker2011,Dussan11}, we allow for different nonlocalities above and below the Fermi energy, otherwise the symmetry around this energy is essentially maintained by the fit.
The values of the nonlocality parameters $\beta$ appear reasonable and range from 0.64 fm above to 0.81 fm below the Fermi energy for volume absorption.
These parameters are critical in ensuring that particle number is adequately described. 
We limit contributions to $\ell \le 5$ below $\varepsilon_F$~\cite{Dussan11} obtaining 19.88 protons and 19.79 neutrons. 
We note the extended energy domain for volume absorption below $\varepsilon_F$ to accommodate the Jefferson Lab data.
Surface absorption requires nonlocalities of 0.94 fm above and 2.07 fm below $\varepsilon_F$~\cite{Hossein13}. 

\begin{figure}[bpt]
\includegraphics*[scale=0.4]{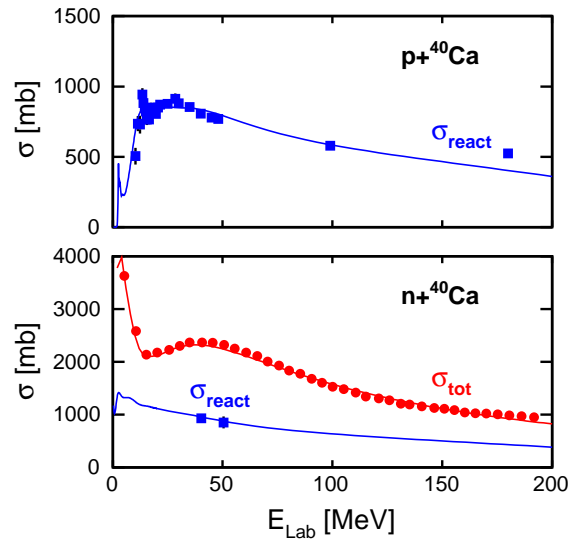}
\caption{(Color online) Total reaction cross sections are displayed as a
function of proton energy while both total and reaction cross sections are shown for neutrons.}
\label{fig:totalc}
\end{figure}
The final fit to the experimental elastic scattering data is shown in Fig.~\ref{fig:elast} 
while the fits to total and reaction cross sections are shown in Fig.~\ref{fig:totalc}.
In all cases, the quality of the fit is the same as in Refs.~\cite{Charity07} or \cite{Mueller11}.
This statement also holds for the analyzing powers given in Ref.~\cite{Hossein13}.
%Total and reaction cross sections for neutrons as well as reaction cross sections for protons are shown in Fig.~%\ref{fig:totalc}.
%Again the quality of the fit is as good as that obtained previously.

\begin{figure}[bp]
\includegraphics*[scale=0.42]{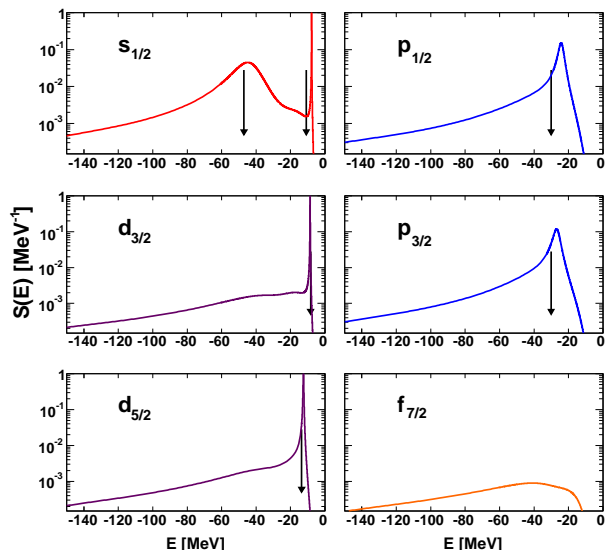}
\caption{(Color online) Spectral strength for protons in the $\ell j$-orbits which are fully occupied in the independent-particle model as well as the $f_{7/2}$ strength associated with the first empty orbit in this description. 
The arrows indicate the experimental location of the valence states as well as the peak energies for the distributions of deeply bound ones.}
\label{fig:spectral}
\end{figure}
Having established our description at positive energies is equivalent to our earlier work, but now consistent with theoretical expectations associated with the nonlocal content of the nucleon self-energy, we turn our attention to the new results below the Fermi energy.
In Fig.~\ref{fig:spectral} we display the spectral strength given in Eq.~(\ref{eq:specs}) as a function of energy for the first few levels in the independent-particle model.
The downward arrows identify the experimental location of the levels near the Fermi energy while
for deeply bound levels they correspond to the peaks obtained from $(p,2p)$~\cite{Jacob73} and $(e,e'p)$ reactions~\cite{FM84}.
The DOM strength distributions track the experimental results represented by their peak location and width.

For the quasi-hole states we find spectroscopic factors of 0.78 for both the $1s_{1/2}$ and 0.76 for the $0d_{3/2}$ level.
The location of the former deviates slightly from the experimental peak which may require 
additional state dependence of the self-energy as expressed by poles nearby in energy~\cite{Dickhoff04}.
%As for a comparison with the spectroscopic factors that were obtained from the $(e,e'p)$ reaction in Ref.~\cite{Kramer89}, 
The analysis of the $(e,e'p)$ reaction in Ref.~\cite{Udias95} clarified that the treatment of nonlocality in the relativistic approach leads to different distorted proton waves as compared to conventional non-relativistic optical potentials, yielding about 10-15\% larger spectroscopic factors.
Our current results are also larger by about 10-15\% than the numbers extracted in Ref.~\cite{Kramer89}.
Introducing local DOM potentials in the analysis of transfer reactions has been shown to have salutary effects for the extraction of spectroscopic information of neutrons~\cite{Nguyen2011} and nonlocal potentials should further improve such analyses. 

\begin{figure}[bpt]
\includegraphics*[scale=0.35]{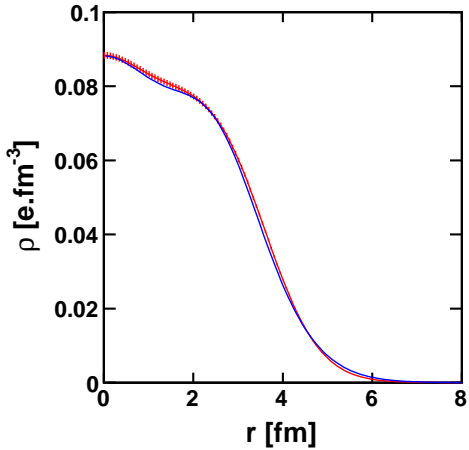}
\caption{(Color online) Comparison of experimental charge density~\protect\cite{deVries1987} (thick line) with the DOM fit (thin line).}
\label{fig:charged}
\end{figure}
In Fig.~\ref{fig:charged} we compare the experimental charge density of ${}^{40}$Ca (thick line representing a 1\% error) with the DOM fit.
While some details could be further improved, it is clear that an excellent description of the charge density is possible in the DOM.
The correct particle number is essential for this result which in turn can only be achieved by including nonlocal absorptive potentials that are also constrained by the high-momentum spectral functions.
With a local absorption we are not capable to either generate a particle number close to 20 or describe the charge density accurately~\cite{Dickhoff10}.

%\begin{figure}[tbp]
%\includegraphics*[scale=0.35]{analyzing}
%\caption{(Color online) Total reaction cross sections are displayed as a
%function of proton energy.}
%label{fig:react}
%\end{figur

\begin{figure}[tbp]
\includegraphics*[scale=0.45]{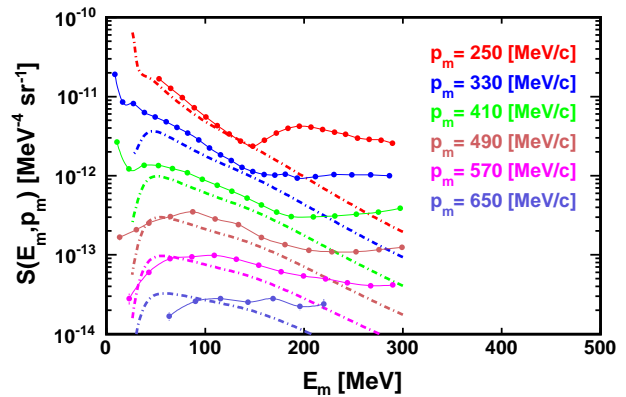}
\caption{(Color online) Spectral strength as a function missing energy for different missing momenta as indicated in the figure. The data  are the average of the ${}^{27}$Al and ${}^{56}$Fe measurements from~\protect\cite{Rohe04A}.}
\label{fig:highk}
\end{figure}

We compare in Fig.~\ref{fig:highk} the results for the high-momentum removal spectral strength with the Jefferson Lab data~\cite{Rohe04A}.
We note that the high-energy data correspond to intrinsic nucleon excitations and cannot be part of the present analysis.
To further improve the description, one would have to introduce an energy dependence of the radial form factors for the potentials. Nevertheless we conclude that an adequate description is generated which corresponds to 10.6\% of the protons having momenta above 1.4 $\textrm{fm}^{-1}$.
Employing the energy sum rule~\cite{Dickhoff08} in the form given in Ref.~\cite{Dieperink74}, 
yields a binding energy of 7.91 MeV/$A$ much closer to the experimental 8.55 MeV/$A$ than found in Ref.~\cite{Dickhoff10}.
The constrained presence of the high-momentum nucleons is responsible for this change~\cite{Muther95}.
The 7.91 MeV/$A$ binding obtained here represents the contribution to the ground-state energy from two-body interactions including a kinetic energy of 22.64 MeV/$A$ and was not part of the fit.
This empirical approach therefore leaves about 0.64 MeV/$A$ attraction for higher-body interactions about 1 MeV/$A$  less than the Green's function Monte Carlo results of Ref.~\cite{Pieper01} for light nuclei.

In conclusion we have demonstrated that the nucleon self-energy for ${}^{40}$Ca requires a nonlocal form and can then with reasonable assumptions represent all relevant single-particle properties of this nucleus.

%\begin{figure}[tbp]
%\includegraphics*[scale=0.3]{n_of_kn}
%\caption{(Color online) Quantities derived from DOM fits. a) The energy
%dependence of the Hartree-Fock potential $V_{HF}$, and the surface $W_{s}$
%and volume $W_{v}$ imaginary potentials. b) The radial dependence of the
%effective mass at $E_{F}$. c) The occupation probabilities (points indicate
%sp levels). For all plots, the thick (back), medium (red) and thin (green)
%curves give the results for $^{40}$Ca, $^{48}$Ca, and $^{60}$Ca,
%respectively.}
%\label{fig:nofk}
%\end{figure}

%In conclusion, the properties of protons and neutrons in ${}^{40}$Ca are accurately described by the DOM self-energy %both at positive energy as well as those associated with ground state properties.

This work was supported by the U.S. Department of Energy,
Division of Nuclear Physics under grant DE-FG02-87ER-40316 and the U.S.
National Science Foundation under grants PHY-0968941 and PHY-1304242.

\bibliography{DOMbib_W}

\end{document}